\documentclass[runninghead]{llncs}
\pdfoutput=1

\usepackage{fancyhdr} 
\usepackage[ruled]{algorithm2e}

\usepackage{perpage}
\MakePerPage{footnote}

\usepackage[flushleft]{threeparttable}
\usepackage{makecell}
\usepackage{comment}

\usepackage{graphicx}
\usepackage{hyperref}
\usepackage{subfiles}
\usepackage{float}

\usepackage{mathtools}

\usepackage{lscape}
\usepackage[figuresright]{rotating}
\usepackage[graphicx]{realboxes}
\usepackage{adjustbox}
\usepackage{caption}
\usepackage{listings}
\usepackage{booktabs}
\usepackage{subfigure}
\usepackage{multirow}

\usepackage{tikz}
\usetikzlibrary{arrows.meta}
\usepackage{framed} 

\usepackage{amsmath}
\usepackage{amssymb}

\usepackage{lscape}
\usepackage{appendix}

\usepackage{comment}

\usepackage{setspace}
\usepackage{ntheorem}

\usepackage{hyperref} 
\makeatletter
\def\UrlAlphabet{%
      \do\a\do\b\do\c\do\d\do\e\do\f\do\g\do\h\do\i\do\j%
      \do\k\do\l\do\m\do\n\do\o\do\p\do\q\do\r\do\s\do\t%
      \do\u\do\v\do\w\do\x\do\y\do\z\do\A\do\B\do\C\do\D%
      \do\E\do\F\do\G\do\H\do\I\do\J\do\K\do\L\do\M\do\N%
      \do\O\do\P\do\Q\do\R\do\S\do\T\do\U\do\V\do\W\do\X%
      \do\Y\do\Z}
\def\UrlDigits{\do\1\do\2\do\3\do\4\do\5\do\6\do\7\do\8\do\9\do\0}
\g@addto@macro{\UrlBreaks}{\UrlOrds}
\g@addto@macro{\UrlBreaks}{\UrlAlphabet}
\g@addto@macro{\UrlBreaks}{\UrlDigits}
\makeatother

\usepackage[
	n,
	operators,
	advantage,
	sets,
	adversary,
	landau,
	probability,
	notions,	
	logic,
	ff,
	mm,
	primitives,
	events,
	complexity,
	asymptotics,
	keys]{cryptocode}
	
\usepackage{dashbox}
\usepackage{tikz}
\usepackage{enumerate}
\usepackage{enumitem}

\usepackage{etoolbox}
\usepackage{tkz-euclide}

\usepackage{booktabs,tabulary,array,parskip}
\usepackage{xcolor}

\usepackage{caption} 
\begin{document}

\title{Non-Fungible Token (NFT): Overview, Evaluation, Opportunities and Challenges}
\subtitle{(Tech Report$^{\textrm{V}2}$)}

\renewcommand\rightmark{}
\renewcommand\leftmark{NFT Overview}
\newcommand{\romannum}[1]{\romannumeral #1}




\author{Qin Wang$^{\star}$\inst{2,4}, Rujia Li$\thanks{These authors contributed equally to the work.\\ Email: \url{qinwang@swin.edu.au} and \url{rxl635@bham.ac.uk}.}$\inst{1,3}, Qi Wang\inst{1}, Shiping Chen\inst{4}}
\authorrunning{Qin Wang}
\authorrunning{Rujia Li}
\titlerunning{NFT Overview}

\institute{
Southern University of Science and Technology
\\
\and
Swinburne University of Technology
\\
\and
University of Birmingham
\\
\and
CSIRO Data61
\\
}
\maketitle           

\begin{abstract}

The Non-Fungible Token (NFT) market is mushrooming in recent years. The concept of NFT originally comes from a token standard of Ethereum, aiming to distinguish each token with distinguishable signs. This type of token can be bound with virtual/digital properties as their unique identifications. With NFTs, all marked properties can be freely traded with customized values according to their ages, rarity, liquidity, etc. It has greatly stimulated the prosperity of the decentralized application (DApp) market. At the time of writing (May 2021), the total money used on completed NFT sales has reached $34,530,649.86$ USD. The thousandfold return on its increasing market draws huge attention worldwide. However, the development of the NFT ecosystem is still in its early stage, and the technologies of NFTs are pre-mature. Newcomers may get lost in their frenetic evolution due to the lack of systematic summaries. In this technical report, we explore the NFT ecosystems in several aspects. We start with an overview of state-of-the-art NFT solutions, then provide their technical components, protocols, standards, and desired proprieties. Afterwards, we give a security evolution, with discussions on the perspectives of their design models, opportunities and challenges. To the best of our knowledge, this is the first systematic study on the current NFT ecosystems.

\end{abstract}

\section{Introduction}

\underline{N}on-\underline{F}ungible \underline{T}oken (NFT) is a type of cryptocurrency \cite{fairfield2021tokenized} that is derived by the smart contracts of Ethereum \cite{wood2014ethereum}. NFT was firstly proposed in Ethereum Improvement Proposals (EIP)-721 \cite{eip721} and further developed in EIP-1155 \cite{eip1155}. NFT differs from classical cryptocurrencies \cite{shirole2020cryptocurrency} such as Bitcoin~\cite{nakamoto2019bitcoin} in their intrinsic features. Bitcoin~\cite{nakamoto2019bitcoin} is a standard coin in which all the coins are equivalent and indistinguishable. In contrast, NFT is unique which cannot be exchanged like-for-like (equivalently, non-fungible), making it suitable for identifying something or someone in a unique way. To be specific, by using NFTs on smart contracts (in Ethereum~\cite{wood2014ethereum}), a creator can easily prove the existence and ownership of digital assets in the form of videos, images, arts \cite{franceschet2020crypto}, event tickets \cite{regner2019nfts}, etc. Furthermore, the creator can also earn royalties each time of a successful trade on any NFT market or by peer-to-peer exchanging. Full-history tradability, deep liquidity, and convenient interoperability enable NFT to become a promising intellectual property (IP)-protection solution. Although, in essence, NFTs represent little more than code, but the codes to a buyer have ascribed \textit{value} when considering its comparative scarcity as a digital object. It well secures selling prices of these IP-related products that may have seemed unthinkable for non-fungible virtual assets.

In recent years, NFTs have garnered remarkable attention from both the industrial and scientific communities. It was reported that the 24-hour trading volume on average of the NFT market is $4,592,146,914$ USD\footnote{Data source from Coingecko (May, 2021) \url{https://www.coingecko.com/en/nft}}, while the 24-hour trading volume of the entire cryptocurrency market is $341,017,001,809$ USD. The liquidity of NFT-related solutions has accounted for 1.3$\%$ of the entire cryptocurrency market in such a short period (5 months). Early investors obtain thousandfold returns by selling unique digital collectibles. At the time of writing (May 2021), the NFTs-related market has significantly increased compared to one year ago (January 2020). Specifically, the total number of sales is $25,729$ and their total amounts spent on completed sales reach $34,530,649.86$ USD\footnote{Data captured from \url{https://nonfungible.com/market/history}}. In particular, the total number of primary-market sales occupies $17,140$, while the number of secondary sales (user-to-user) is $8,589$. Correspondingly, the total USD used on primary market sales is $8,816,531.10$. Besides, the active market wallets achieve $12,836$, which is still increasing at a high speed as time goes. Surprisingly, the sale of NFTs was estimated at $12$ million (December 2020) but exploded to $340$ million within just two months (February 2021). Such skyrocketing development makes NFT become a craze, or even be described by some as the future of digital assets. 

Besides the above data, people have expressed interest in various types of NFTs. They participate in NFT-related games or trades with enthusiasm. CryptoPunks \cite{cryptopunks}, one of the first NFT on Ethereum, has created more than $10,000$ collectible punks ($6039$ males and $3840$ females) and further promoted the ERC-721 standard to become popular. CryptoKitties \cite{cryptokitties} officially put NFTs on notice, and hit the market in 2017 with the gamification of the breeding mechanics. Participants fiercely competed at high prices to auction the rare cats, and the highest price reaches more than 999 ETH\footnote{Price available: \url{https://www.cryptokitties.co/kitty/1866420}} (equally 3M USD). Another outstanding instance is NBA Top Shot \cite{nbatopshot}, which is an NFT trading platform used to buy/sell digital short videos of NBA moments. Thousands of NBA fans from around worldwide have collected over $7.6$ million top shot moments, building the roster of rookies, vets, and rising star players. Following projects also enjoy great success including Picasso Punks \cite{picassopunks}, Hashmasks \cite{hashmasks}, 3DPunks \cite{3dpunks},  unofficial punks \cite{unofficialpunks}, Polkamon \cite{polkamon}, Chubbies \cite{chubbies}, Bullrun Babes \cite{bullrun}, Aavegotchi \cite{aavegotchi}, CryptoCats \cite{cryptocats}, Moon Cats Rescue \cite{mooncatsrescue}, NFT box \cite{nftbox}, etc. There is no doubt that there is a hype cycle surrounding NFTs where most of products can be sold with high prices, some even hundreds or thousands of ETHs. Besides games and collectibles, NFTs also promote the development of art \cite{trautman2021virtual}, ticketing event \cite{regner2019nfts}, value \cite{chevet2018blockchain},\cite{chohan2021non}, IoT \cite{omar2020capability}, and finance \cite{dowling2021fertile}\cite{musan2020nft}. Other types of surrounding markets play important roles as well to provide instant information and secure environments like statistic websites (e.g. NonFungible \cite{NonFungible}, DappRadar \cite{dappradar}, NFT bank\cite{nftbank}, DefiPulse \cite{defipulse}, Coingecko \cite{coingecko}), trading marketplace (cryptoslam \cite{cryptoslam}, Opensea \cite{opensea}, SuperRare \cite{superrare}, Nifty Gateway \cite{niftygateway}, Rarible \cite{rarible}, Zora \cite{zora}) and so-called NFT ecosystem (such as Dego \cite{dego}).

Despite NFTs have a tremendous potential impact on the current decentralized markets and future business opportunities, the NFT technologies are still in the very early stage. Some potential challenges are required to be carefully tackled, while some promising opportunities should be highlighted. Further, even though much literature on NFTs, from blogs, wikis, forum posts, codes and other sources, are available to the public, a systematic study is absent. This paper aims to draw attention to these questions insofar as observed and focus on summarising current NFT solutions. We provide a detailed analysis of its core components, technology roadmap status, opportunities and challenges. The contributions are provided as follows.

\begin{itemize}
    \item[-] Firstly, \textit{we abstract the design models of current NFT solutions.} Specifically, we identify the core technical components that are used to construct NFTs. Then, we present their protocols, standards and targeted properties. 
    
    \item[-] Secondly, \textit{we give a security evaluation of current NFT systems.} We adopt the STRIDE threat and risk evaluation~\cite{shostack2008experiences} to investigate potential security issues. Based on that, we also discuss the corresponding defense measures for the issues.
    
    \item[-] Thirdly, \textit{we explore some future opportunities of NFTs in many fields.} Applying NFTs to real scenarios will boost a wide range of new applications. We give a set of practical instances (projects) leveraging NFTs with great success or prosperous markets.
    
    \item[-] Finally, \textit{we highlight a series of open challenges in NFT ecosystems.} Blockchain-based NFT systems still confront unavoidable problems like privacy issues, data inaccessibility, etc. We outline open challenges existed in state-of-the-art NFT solutions.

\end{itemize}
The rest part of this work is organized as follows. Section \ref{sec-model} provides the technical components used to build NFTs. Section \ref{sec-proto} presents the protocols and standards. Based on that, Section \ref{sec-security} gives our security evaluation. Section \ref{sec-oppo} discusses the future opportunities, while Section \ref{sec-chanllen} outlines the open challenges. Finally, Section \ref{sec-conclu} concludes this work. Appendix A gives our version updates. Appendix B and Appendix C provide NFT collectible ranking and an overview of existing NFT projects, respectively. Appendix D presents a detailed instance analysis. 

\section{Technical Components}
\label{sec-model}
In this part, we show technical components related to the NFT's activities. These components lay the building foundations of a fully functional NFT scheme. 

\smallskip
\noindent\textbf{Blockchain.} Blockchain was originally proposed by Nakamoto~\cite{nakamoto2019bitcoin}, where Bitcoin uses the proof of work (PoW) \cite {gervais2016security} algorithm to reach an agreement on transaction data in a decentralized network. Blockchain is defined as a distributed and attached-only database that maintains a list of data records linked and protected using cryptographic protocols~\cite{garay2017bitcoin}. Blockchain provides a solution to the long-standing Byzantine problem \cite{lamport2019byzantine}, which has been agreed upon with a large network of untrusted participants. Once the shared data on the blockchain is confirmed in most distributed nodes, it becomes immutable because any changes in the stored data will invalidate all subsequent data. The most prevailing blockchain platform used in NFT schemes is Ethereum \cite{wood2014ethereum}, providing a secure environment for executing the smart contracts. In addition, several solutions drop their customized chain-engines or blockchain platforms to support their specialized applications, and some of them are Flow \cite{flow2020}, EOS \cite{wax}, Hyperledger \cite{hong2019design}\cite{bal2019nftracer}, and Fast Box \cite{fastbox}\cite{wang2021weak}.

\smallskip
\noindent\textbf{Smart Contract.} Smart contracts were originally introduced by Szabo \cite {szabo1996smart}, aiming to accelerate, verify or execute digital negotiation. Ethereum~\cite {wood2014ethereum} further developed smart contracts in the blockchain system~\cite {garay2020sok}\cite{bano2019sok}. Blockchain-based smart contracts adopt Turing-complete scripting languages to achieve complicated functionalities and execute thorough state transition replication over consensus algorithms to realize final consistency. Smart contracts enable unfamiliar parties and decentralized participants to conduct fair exchanges without a trusted third party and further propose a unified method to build applications across a wide range of industries. The applications operating on top of smart contracts are based on state-transition mechanisms. The states that contain the instructions and parameters are shared by all the participants, thus guaranteeing transparency of the execution of these instructions. Also, the positions between states have to stay the same across distributed nodes, which is important to its consistency. Most NFT solutions  \cite{cryptopunks}\cite{cryptokitties}\cite{nbatopshot}\cite{picassopunks}\cite{hashmasks}\cite{3dpunks} rely on smart contract-based blockchain platforms to ensure their order-sensitive executions.

\smallskip
\noindent\textbf{Address and Transaction.} Blockchain address and transaction are the essential concepts in cryptocurrencies. A blockchain address is a unique identifier for a user to send and receive the assets, which is similar to a bank account when spending the assets in the bank. It consists of a fixed number of alphanumeric characters generated from a pair of public key and private key. To transfer NFTs, the owner must prove in possession of the corresponding private key and send the assets to another address(es) with a correct digital signature. This simple operation is usually performed using a cryptocurrency wallet and is represented as sending a transaction to involve smart contracts in the ERC-777 \cite{erc777} standard.

\smallskip
\noindent\textbf{Data Encoding.} Encoding is the process of converting data from one form to another. Normally, many files are often encoded into either efficient, compressed formats for saving disk space or into an uncompressed format for high quality/resolution. In the mainstream blockchain systems such as Bitcoin \cite{nakamoto2019bitcoin} and Ethereum \cite{wood2014ethereum}, they employ \texttt{hex} values to encode transaction elements such as the function names, parameters and return values. This implies that the raw NFT data must follow these rules. If one claims s/he owns the NFT-based intellectual property, s/he essentially owns the original piece of \texttt{hex} values signed by the creator. Others can freely copy the raw data, but they cannot claim ownership of the property. Based on that, we can observe that the NFT-related activities (e.g. buy/sell/trade/auction) have to be processed under these four phases, similar to the basic processing procedure of smart contracts.

\section{Protocols, Standards and Properties}
\label{sec-proto}

This section presents two basic models of NFT schemes, with emphasis on their protocols, token standards and key properties.


\subsection{Protocols}

\begin{figure}[!htb]
    \centering
    \includegraphics[width=0.9\linewidth]{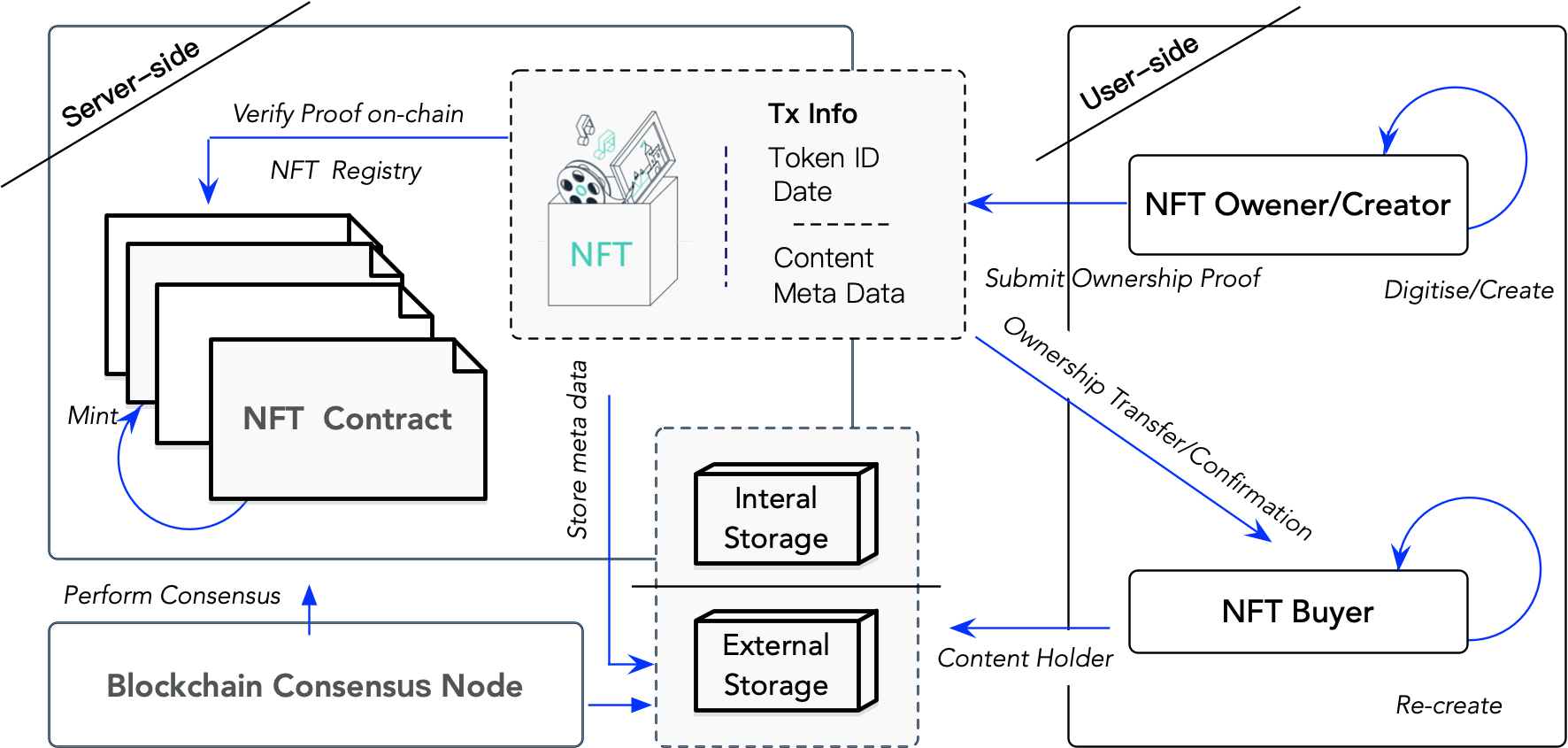}
    \caption{Workflow of NFT Systems}
    \label{fig-erc1}
\end{figure}

The establishment of NFT requires an underlying distributed ledger for records, together with exchangeable transactions for trading in the peer-to-peer network. This report primarily treats the distributed ledger as a special type of database to store NFT data. In particular, we assume that the ledger has basic security consistency, completeness, and availability characteristics. Based on that, we identify two design patterns for the NFT paradigm. The former protocol is established from top to bottom with a very simple yet classical path: \textit{building NFTs from the initiator, and then sell them to the buyer}. In contrast, the later route (e.g. Loot \cite{loot}, detailed analysis in Appendix D) reverses this path: \textit{setting a NFT template, and every user can create their unique ontop NFTs.} We separately provide detailed protocols of these two design patterns as below. To be noted, for both of them, they still follow a very similar workflow when executed on blockchain systems (cf. Fig.\ref{fig-erc1}), meaning that different designs will not change the underlying operating mechanism.

\textbf{Top to Bottom.} 
For the first design (e.g., CryptoPunks \cite{cryptopunks}), an NFT protocol consists of another two roles: NFT owner and NFT buyer. 

\begin{itemize}

\item[-] \textbf{NFT Digitize.} An NFT owner checks that the file, title, description are completely accurate. Then, s/he digitizes the raw data into a proper format.

\item[-] \textbf{NFT Store.} An NFT owner stores the raw data into an external database outside the blockchain. Note that, s/he is also allowed to store the raw data inside a blockchain, despite this operation is gas-consuming.
 
\item[-] \textbf{NFT Sign.} The NFT owner signs a transaction, including the hash of NFT data, and then sends the transaction to a smart contract.

\item[-] \textbf{NFT Mint\&Trade.} After the smart contract receives the transaction with the NFT data, the minting and trading process begins. The main mechanism behind NFTs is the logic of the Token Standards that can be found in Section~\ref{sec:token}. 


\item[-] \textbf{NFT Confirm.} Once the transaction is confirmed, the minting process completes. By this approach, NFTs will forever link to a unique blockchain address as their persistence evidence.

\end{itemize}

\textbf{Bottom to Top.}
For this design (e.g., Loot \cite{loot}), the protocol consists of two roles: NFT creator and NFT buyer. In most cases, a buyer can also act as a creator because an NFT product is created based on random seeds when a buyer bids for it. This extends the functions in terms of user customization. Here, we use the superscript $*$ to highlight differences compared with the previous one.

\begin{itemize}
\item[-] \textbf{Template Create$^*$.} The project founder initiates a template via the smart contract to set up several basic rules, such as different features (character style, weapons, or accessories) in the game. 

\item[-] \textbf{NFT Randomize$^*$.} Once a buyer bids for an NFT, s/he can customize the NFT product with a set of additional features on top of basic lines. These additional features are randomly selected from a database that was predefined at the initial state.  

\item[-] \textbf{NFT Mint\&Trade.} The minting and trading process starts once the corresponding smart contract is triggered.

\item[-] \textbf{NFT Confirm.} All the procedures are conducted through smart contracts. The generated NFT will be persistently stored on-chain when the consensus procedure has been completed. 
\end{itemize}

In a blockchain system, each block has a limited capacity. When the capacity in one block becomes full, other transactions will enter a future block linked to the original data block. In the end, all linked blocks have created a long-term history that remains permanent. The NFT system, in essence, is a blockchain-based application. Whenever an NFT is minted or sold, a new transaction is required to send to invoke the smart contract. After the transaction is confirmed, the NFT metadata and ownership details are added to a new block, thereby ensuring that the history of the NFT remains unchanged and the ownership is preserved.

\subsection{Token Standards}
\label{sec:token}
In this part, we clarify token standards related to NFTs, including ERC-20 \cite{erc20}, ERC-721 \cite{erc721}, and ERC-1155 \cite{erc1155} (see Algorithm \ref{algorithm1}). These standards have a great impact on the ongoing NFT schemes. We discuss them as follows.

\begin{algorithm} 
\caption{NFT Standard Interfaces (with selected functions) }\label{algorithm1}
\BlankLine
 \textbf{interface} ERC721 \{ \\
 \quad    function \texttt{ownerOf}(uint256$\_$tokenId) external view returns (address); \\
 \quad    function \texttt{transferFrom}(address$\_$from, address$\_$to, uint256$\_$tokenId) external payable;  ...\\
      \}    \\
 \textbf{interface} ERC1155 \{ \\
 \quad    function \texttt{balanceOf}(address$\_$owner, uint256$\_$id) external view returns (address);  \\
 \quad    function \texttt{balanceOfBatch}(address calldata $\_$owners, uint256 calldata $\_$ids) external view returns (uint256 memory);\\
 \quad    function \texttt{transferFrom}(address$\_$from, address$\_$to, uint256$\_$id, uint256 quantity)   external payable; ...\\
      \} \\
\end{algorithm}

The most prevailing token standard comes from ERC-20 \cite{erc20}. It introduces the concept of fungible tokens that can be issued on top of Ethereum once satisfying the requirements. The standard makes tokens the same as another one (in terms of both type and value). An arbitrary token is always equal to all the other tokens. This stimulates the hype of Initial Coin Offering (ICO) from 2015 to present. A lot of public chains and various blockchain-based DApps  \cite{cai2018decentralized}\cite{raval2016decentralized} gain sufficient initial fundings in this way. In contrast, ERC-721 \cite{erc721} introduces a non-fungible token standard that differs from the fungible token. This type of token is unique that can be distinguished from another token. Specifically, every NFT has a uint256 variable called \texttt{tokenId}, and the pair of contract address and uint256 \texttt{tokenId} is globally unique. Further, the \texttt{tokenId} can be used as an input to generate special identifications such as images in the form of zombies or cartoon characters.

\begin{figure}[!htb]
    \centering
    \includegraphics[width=0.95\linewidth]{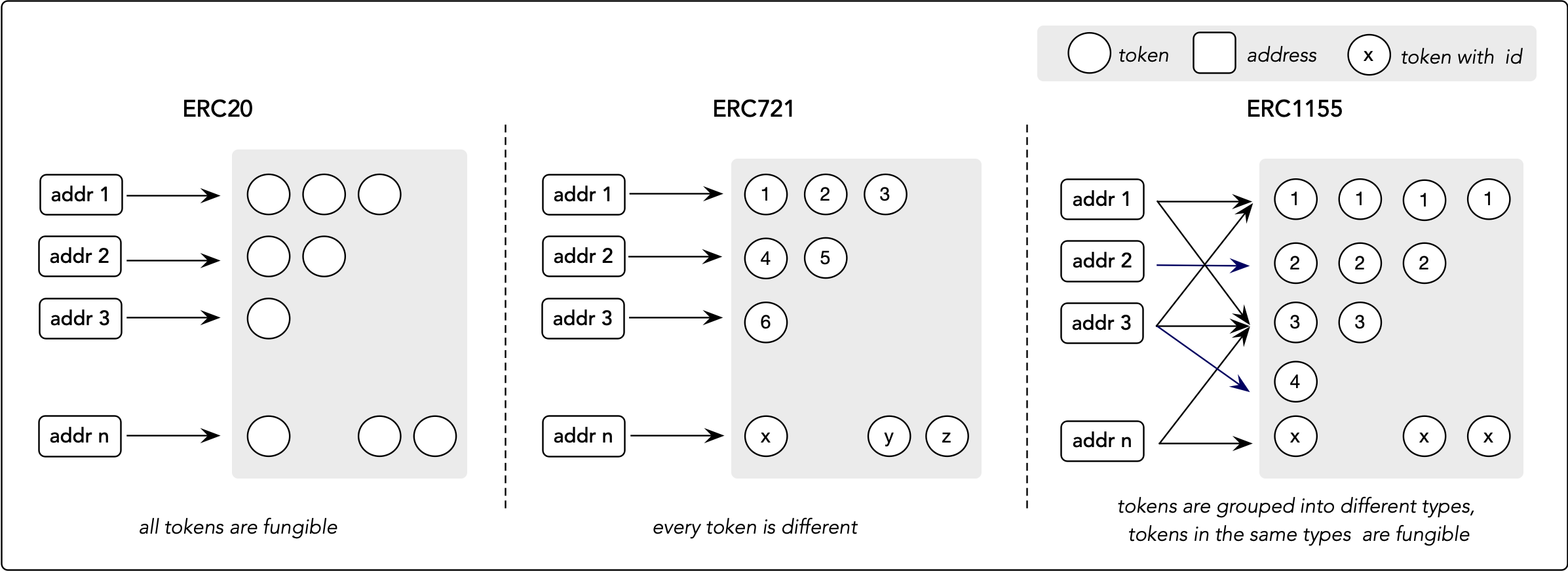}
    \caption{NFT-related Token Standards}
    \label{fig-erc2}
\end{figure}

Another standard ERC-1155 (Multi Token Standard) \cite{erc1155} extends the representation of both fungible and non-fungible tokens. It provides an interface that can represent any number of tokens. In previous standards, every \texttt{tokenId} in contact only contains a single type of tokens. For instance, ERC-20 makes each token type deployed in separate contracts. As well, ERC-721 deploys the group of non-fungible tokens in a single contract with the same configurations. In contrast, ERC-1155 extends the functionality of \texttt{tokenId}, where each of them can independently represent different configurable token types. The field may contain its customized information such as the metadata, lock-time, date, supply, or any other attributes. Here, we provide an illustration (see Fig.\ref{fig-erc2}) to show their structures and aforementioned differences.

\subsection{NFTs Desired Proprieties}

NFT schemes are essentially decentralized applications~\cite{buterin2014next}, and thus enjoy the benefits/properties from their underlying public ledgers. We summarise the key properties as follows.

\begin{itemize}

\item[-] \textbf{Verifiability.} The NFT with its token metadata and its ownership can be publicly verified.

\item[-] \textbf{Transparent Execution.} The activities of NFTs include minting, selling and purchasing are publicly accessible.

\item[-] \textbf{Availability.} The NFT system never goes down. Alternatively, all the tokens and issued NFTs are always available to sell and buy.

\item[-] \textbf{Tamper-resistance.} The NFT metadata and its trading records are persistently stored and cannot be manipulated once the transactions are deemed as confirmed.

\item[-] \textbf{Usability.} Every NFT has the most up-to-date ownership information, which is user-friendly and information-clearly. 

\item[-] \textbf{Atomicity.} Trading NFTs can be completed in one atomic, consistent, isolated, and durable (ACID) transaction. The NFTs can run in the same shared execution state. 


\item[-] \textbf{Tradability.} Every NFTs and its corresponding products can be arbitrarily traded and exchanged.


\end{itemize}

\section{Security Evaluation}
\label{sec-security}

An NFT system is a combination technology that consists of blockchain, storage and web application. Security evaluation on the NFT system is challenging since each component may become an attacking interface that makes the whole system really vulnerable against the attacker. Thus, we adopt the STRIDE threat and risk evaluation~\cite{shostack2008experiences}, which covers all security aspects of a system: authenticity, integrity, non-repudiability, availability and access control. We investigate the potential security issues and propose some of the corresponding defense measures to address these issues (see Table~\ref{tab-stride}).

\textbf{Spoofing.} Spoofing is the ability to impersonate another entity (for example, another person or computer) on the system, which corresponds to authenticity. When a user interacts to mint or sells NFTs, a malicious attacker may exploit authentication vulnerabilities or steal the user's private key to transfer the ownership of NFTs illegally. Thus, we recommend having a formal verification for the NFT smart contract and to use the cold wallet to prevent private key leakage.

\textbf{Tampering.} Tampering refers to the malicious modification of NFT
data, which violates integrity. Assume that the blockchain is a \textit{robust} public transaction ledger~\cite{garay2015bitcoin,garay2017bitcoin} and a hash algorithm is preimage resistance and second preimage resistance~\cite{rogaway2004cryptographic}. The metadata and ownership of NFTs cannot be maliciously modified after the transaction is confirmed. However, the data stored outside blockchain may be manipulated. Therefore, we recommend users to send both the hash data as well as the original data to the NFT buyer when trading/exchanging NFT-related properties.

\textbf{Repudiation.} Repudiation refers to the situation where the author of a statement cannot dispute~\cite{zhou1996fair}, which is related to the security property of non-repudiability~\cite{menezes2018handbook}. In particular, the fact that a user sends NFT to another user cannot deny. This is guaranteed by the security of the blockchain and the unforgeability property of a signature scheme. However, the hash data may be tampered by a malicious attacker, or the hash data may bind with an attacker's address. Thus, we believe that using a multi-signature contract can partly solve this issue since each binding must be confirmed by more than one participant. 

\textbf{Information Disclosure.} Information leakage occurs when information is exposed to unauthorized users, which violates confidentiality~\cite{menezes2018handbook}. In the NFT system, the state information and the instruction code in the smart contracts are entirely transparent, and any state and its changes are publicly accessible by any observer. Even if the user only puts the NFT hash into the blockchain, the malicious attackers can easily exploit the linkability of the hash and transaction. Thus, we recommend the NFT developer to use privacy-preserving smart contracts~\cite{li2019auditable}\cite{li2020accountable} instead of plain smart contracts to protect the user's privacy.

\begin{table}
\def\item{\hangindent1em\textbullet~}
\tymin=0pt
\tymax=80pt
\caption{Potential Security Issues and Corresponding Solutions of NFTs.}
\label{tab-stride}
\resizebox{\linewidth}{!}{ 
\begin{tabulary}{\linewidth}{C|p{5cm}|p{5cm}}

\toprule
   \textbf{STRIDE} & \quad\quad\quad\quad Security Issues & \quad\quad\quad\quad\quad Solutions \\\midrule
 
\textbf{S}poofing \textit{(Authenticity)}
& 
\quad \item An attacker may exploit authentication vulnerabilities\par\smallskip
\quad \item An attacker may steal a user's private key.
& 
\quad \item A formal verification on the smart contract.\par\smallskip
\quad \item Using the cold wallet to prevent the private key leakage.
\\ 
\midrule

\textbf{T}ampering \textit{(Integrity)} 
& 
\quad \item The data stored outside the blockchain may be manipulated.
& 
\quad \item 
Sending both the original data and hash data to the NFT buyer when trading NFTs.
\\ 
\midrule

\textbf{R}epudiation \textit{(Non-repudiability)} 
& 
\quad \item The hash data may bind with an attacker's address.
& 
\quad \item Using a multi-signature contract partly.

\\ 
\midrule

\textbf{I}nformation disclosure \textit{(Confidentiality)} 
& 
\quad \item An attacker can easily exploit the hash and transaction to link a particular NFT buyer or seller. 
& \quad \item Using privacy-preserving smart contracts instead of smart contracts to protect the user's privacy.
\\ 
\midrule

\textbf{D}enial of service \textit{(Availability)} 
& 
\quad \item The NFT data may become unavailable if the asset is stored outside the blockchain. 
& 
\quad \item Using the hybrid blockchain architecture with 
weak consensus algorithm.

\\
\midrule

\textbf{E}levation of privilege \textit{(Authorization)} 
& 
\quad \item A poorly designed smart contract may make NFTs lose such properties. 
& 
\quad \item A formal verification on the smart contracts.

\\ 

\bottomrule
\end{tabulary}
}
\end{table}

\textbf{Denial of Service (DoS).} DoS attack~\cite{Moore2006InferringID} is a type of network attack in which a malicious attacker aims to render a server unavailable to its intended users by interrupting the normal functions. DoS violates the availability and breaks down the NFT service, which can indeed be used by unauthorized users. Fortunately, the blockchain guarantees the high availability of user's operations. Legitimate users can use the required information when needed and will not lose data resources due to accidental errors. However, DoS can also be used to attack the centralized web applications or the raw data outside the blockchain, resulting in denial-of-service to NFT service. Recently, a new hybrid blockchain architecture with 
weak consensus algorithm was proposed \cite{wang2021weak}, by which this architecture solves the availability issues using two algorithms.


\textbf{Elevation of Privilege.} Elevation of Privilege~\cite{shostack2008experiences} is a property that is related to the authorization. In this type of threat, an attacker may gain permissions beyond those initially granted. In the NFT system, the selling permissions are managed by a smart contract. Again, a poorly designed smart contract may make NFTs lose such properties.

\section{Opportunities}
\label{sec-oppo}

This section explores the opportunities of NFTs. We discuss several typical fields which may get benefits from NFTs.



\smallskip
\noindent\textbf{Boosting Gaming Industry.} 
NFT has great potential in the gaming industry. There already exist some crypto games are CrytpoKitties \cite{cryptokitties}, Cryptocats \cite{cryptocats}, CryptoPunks \cite{cryptopunks}, Meebits \cite{meebits}, Axie Infinity \cite{axieinfinity}, Gods Unchanged \cite{godsunchained}, and TradeStars \cite{tradestars}. A fascinating feature of such games is the ``breeding'' mechanism. Users can personally raise pets and spend much time breeding new offspring. They can also purchase the limited/rare edition virtual pets, and then sell them at a high price. The extra reward attracts lots of investors to join the games, making NFTs come to prominence. Another exciting functions of the NFT are that it provides ownership records of items in the games and promotes economic marking place in the ecosystem, benefiting both developers and players. In particular, game developers who are NFT publishers of the features (e.g., weapons and skins) can earn royalties each time their items are (re-)sold on the open market. The players can obtain personal exclusivity game items. This will create a mutually beneficial business model in which both players and developers profit from the secondary NFT market.  After that, blockchain communities extend NFTs to a large extent that covers various types of digital assets.

\smallskip
\noindent\textbf{Flourishing Virtual Events.} 
Traditional online events rely on centralized companies that provide trust and technology. Although blockchain takes over several types of activities like raising money (either by ICO/IFO/IEO/etc.), its applications are still constrained in a small range of events. NFTs greatly extend the scope of blockchain applications with the help of their additional properties (uniqueness, ownership, liquidity). This enables each individual to link to a specific event just like the patterns in our real life. We give the instance of the ticketing event. When buying tickets in a traditional event ticket market, consumers must trust the third party. Therefore, there is a risk of buying fraudulent or invalid tickets, which are possibly counterfeit or might be cancelled. The same ticket may be sold many times or obtained by extracting from ticket images posted online in an extreme case. ``NFT-based ticket" represents a ticket issued by the blockchain to demonstrate entitlement to access to any event such as culture or sports. An NFT-based ticket is unique and scarce, meaning that the ticket holder cannot resell the ticket after it is sold. The blockchain-based smart contract provides a transparent ticket trading platform for the stakeholders such as the event organizer and the customer. Consumers can buy and sell the crypto ticket from the smart contract rather than rely on third parties in an efficient and reliable way.

\smallskip
\noindent\textbf{Protecting Digital Collectibles.} Digital collectibles contain a variety of types, ranging from trading cards, wines \cite{wiv}, digital images \cite{superrare}, videos \cite{nbatopshot}, virtual real estate \cite{decentraland2020}, domain names \cite{ens}\cite{unstoppabledomains}, diamonds \cite{icecap}\cite{nftdiamonds}, crypto stamps \cite{cryptostamp} and other real/intellectual properties. We take the field of arts as an example. Firstly, artists in traditional ways have very few channels to display the works. The prices cannot reflect the true value of their works due to the absence of attention. Even worse, their published work on social networks has been charged with intermediary fees by platforms and advertisements. NFTs transform their work into digital formats with integrated identities. Artists do not have to transfer ownership and contents to agents. This provides them impetus with lots of profits. Typical instances include Mad Dog Jones's REPLICATOR (selling with 4.1 million USD \cite{maddog}), Grimes's work (selling in total around 6 million USD \cite{grimes}) and other works from great crypto-artists like Beeple \cite{beeple} / Trevor Jones \cite{trevorjonesart}. Furthermore, artists in general cases cannot receive royalties from future sales of their works. In contrast, NFTs can be programmed so that the artist receives a predetermined royalty fee each time when his digital artwork exchanges in the markets (e.g. \cite{knownorigin}, SuperRare \cite{superrare}, MakersPlace \cite{makersplace}, Rare Art Lab\cite{lagelnd}, VIV3 \cite{viv3}). This is an efficient way to manage and protect digital masterpieces. In addition, several platforms (e.g. Mintbase \cite{mintbase}, Mintable \cite{mintable}) have even established tools to support ordinary people to create their own NFT works easily.

\smallskip   
\noindent\textbf{Inspiring the Metaverse.} Metaverse is a collective virtual shared space that allows all types of digital activities. Generally, it covers a set of techniques like augmented reality and the Internet to establish the virtual world. The concept stems from the last decades and has a great progress with the rapid development of blockchain. Blockchain provides an ideal decentralized environment for the virtual online world. Participants under this blockchain-fueled alternative realities \cite{opensea} can have many types of intriguing use cases like enjoying games, displaying self-made arts, trading assets and virtual properties (arts, land parcels, names, video shots, wearables), etc. In addition, users also have opportunities to get profits from the virtual economy. They can lease the buildings (such as offices) to others to earn the bond or raise rare pets and sell them to get the rewards. Primary blockchain-empowered projects are Decentraland \cite{decentraland2020}, Cryptovoxels \cite{cryptovoxels}, Somnium Space \cite{somniumspace}, MegaCryptoPolis \cite{megacryptopolis} and Sandbox \cite{sandbox2020}. In fact, the metaverse ecosystem covers all aforementioned applications. We list it separately here simply because it is still in an early stage due to the complexity.




\section{Challenges}
\label{sec-chanllen}
To enable the development of the above NFT applications, a series of barriers have to be overcome as with any nascent technologies. We discuss some typical challenges from the perspectives of usability, security, governance, and extensibility, covering both the system level issues caused by blockchain-based platforms and human factors such as governs, regulation, and society.  

\subsection{Usability Challenges}
Usability is to measure the users' effectiveness, efficiency, and satisfaction when testing a specific product/design. Most NFT schemes are built on top of Ethereum. Therefore, it is obvious that the main drawbacks of Ethereum still exist. We discuss two major challenges that have direct impacts on the user experience.

\begin{itemize}

\item[-] \textbf{Slow Confirmation.} NFT-related procedures are tyically conducted by sending transactions via the smart contract for reliable and transparent management (such as mint, sell, exchange). However, current NFT systems are closely coupled with their underlying blockchain platforms, which makes them suffer from low performance (Bitcoin reaches merely 7 TPS~\cite{valdeolmillos2019blockchain} while Ethereum only 30 TPS). This results in extremely slow confirmation of NFTs. Conquering this issue requires a redesign of blockchain systems \cite{wang2020sok}, optimization of its structure \cite{wang2019sok}\cite{gudgeon2020sok} or improvement on the consensus mechanisms \cite{bano2019sok}. Existing blockchain systems cannot fulfil such requirements.

\item[-] \textbf{High Gas Prices.} High gas prices have become a major problem for NFT marketplaces, especially when minting the NFTs at a large scale that requires uploading the metadata to the blockchain network. Every NFT-related transactions are more expensive than a simple transfer transaction because smart contracts involve computational resources and storage to be processed. At the time of writing, to mine an NFT token costs over USD $60$ (equivalently in around $ 5 \times 10^{2}wei$) \footnote{Source calculated from \url{https://coinmarketcap.com/} and \url{https://ethereum.org/en/developers/docs/gas/}}. To complete a simple NFT trade can run between USD $60$ and USD $100$ for each transaction. Expensive fees caused by complex operations and high congestion greatly limit its wide adoption. 

\end{itemize}

\subsection{Security and Privacy Issues}
The security of user data pose the first priority of systems. However, the data (stored off-chain but relates to on-chain tags) confronts the risk of losing linkage or being misused by malicious parties. We give details as follows.

\begin{itemize}
\item[-] \textbf{NFT Data Inaccessibility.} In the mainstream NFT projects \cite{flow2020,decentraland2020,sandbox2020}, a cryptographic ``hash'' as the identifier, instead of a copy of the file, will be tagged with the token and then recorded on the blockchain to save the gas consumption. This makes the user lose confidence in the NFT because the original file might be lost or damaged. Several NFT projects integrate their system with a specialized file storage system such as IPFS~\cite{benet2014ipfs} in which IPFS addresses allow users to find a piece of content so long as someone somewhere on the IPFS network is hosting it. Inevitably, such systems have flaws. When the users ``upload'' NFT metadata to IPFS nodes, there is no guarantee that their data will be replicated among all the nodes. The data may become unavailable if the asset is stored on IPFS and the only node storing it is disconnected from the network~\cite{YesYourN79}. This issue has been reported by DECRYPT.IO~\cite{YesYourN79} and CHECKMYNFT.COM~\cite{CH21}. Also, an NFT might point to an erroneous file address. If that is the case, a user cannot prove that s/he actually owns the NFT. In a word, relying on an external system as the core component (storage) for an NFT system is vulnerable.


\item[-] \textbf{Anonymity/Privacy.}
In the current stage, the anonymity and privacy of NFTs are still understudied. Most NFT transactions rely on their underlying Ethereum platform, which only provides pseudo-anonymity rather than strict anonymity or privacy. Users can partially hide their identities if the links between their real identities and corresponding addresses are unknown by the public. Otherwise, all the activities of users under the exposed address are observable. Existing privacy-preserving solutions (e.g. homomorphic encryption \cite{wang2020preserving}, zero-knowledge proof \cite{wang2018designing}, ring signature \cite{noether2015ring}, multi-party computation \cite{raman2018trusted}) have not been yet applied to the NFT-related schemes due to their complicated cryptographic primitives and security assumptions. Similar to other types of blockchain-based systems, decreasing expensive computation costs becomes the key to implement privacy-promised schemes.
\end{itemize}

\subsection{Governance Consideration}

Similar to the situations of most cryptocurrencies, NFTs also confront the barriers like strict management from the government. On the other side, how to properly regulate this nascent technology with the corresponding market is also a challenge. We discuss two typical issues from both sides.

\begin{itemize}

\item[-] \textbf{Legal Pitfalls.} 
NFTs confront legal and policy issues across a wide range of areas \cite{fairfield2021tokenized}\cite{johnson2021decentralized}. Potential concerned areas cover commodities, cross-border transactions, KYC (Know Your Customer) data, etc. It is important to understand the related regulatory scrutiny and litigation before moving into the NFT tracks. In some countries, such as Indian and China, the legal situation is strict for cryptocurrencies, and also for NFT sales. Exchanging, trading, selling, or buying NFTs have to overcome the difficulties of governance. Legally, users can only trade derivates on authorized exchanges such as stocks and commodities or exchange tokens with someone person-to-person. Several countries, such as Malta and France, are trying to implement suitable laws with the aim to regulate the service of digital assets. Elsewhere, issues are resolved by using existing laws. They require buyers to follow complex or even contradictory terms. Therefore, undertaking due diligence is a necessity before investing serious tokens in NFTs.

\item[-] \textbf{Taxable property Issues.}
IP-related products (including arts, books, domain names, etc.) are treated as taxable property under the current legal framework. However, NFT-based sales stay out of this scope. Although few countries, such as the U.S.  (internal revenue service, IRS), tax cryptocurrencies as property, most areas worldwide have not yet considered it. This may greatly increase the financial crimes under cover of NFT trading. The governments would love to make the sale of NFTs reliable with tax consequences. Specifically, the individual participants should have the tax liability on any capital gains that are related to NFT properties. Also, NFT-for-NFT, NFT-for-IP, and Eth-for-NFT (or vice versa) exchanges should be taxed. Furthermore, for high-profit properties, or collectibles, a higher tax bracket should be applied.  Thus,  NFT-related trades are suggested to seek more advice from professional tax departments after the profound discussions.

\end{itemize}

\subsection{Extensibility Issues}
The extensibility of NFT schemes is two-fold. The first is to stress whether a system can interact with other ecosystems. The second focuses on whether NFT systems can obtain updates when the current version is left behind.

\begin{itemize}

\item[-] \textbf{NFT Interoperability (cross-chain).} 
Existing NFT ecosystems are isolated from each other. Users once have selected one type of product can only sell/buy/trade them within the same ecosystem/network. This is due to the reason of its underlying blockchain platform. Interoperability and cross-chain communication are always the handicaps for the wide adoption of DApps. Based on the observations from \cite{zamyatin2019sok}, cross-chain communications can only be implemented with the help of external trusted parties. The decentralization property, in this way, has been inevitably lost to some extent. But fortunately, most of the NFT-related projects adopt Ethereum as their underlying platform. This indicates that they share a similar data structure and can exchange under the same rules. 

\item[-] \textbf{Updatable NFTs.} 
Transitional blockchains update their protocols through soft forks (minor modifications that are compatible forwards) and hard forks (significant modifications that may conflict with previous protocols). A formal discussion has been provided in \cite{ciampi2020updatable} stating the difficulties and trade-offs when applying the updates to an existing blockchain. Despite using the generic model, a new version still faces strict requirements such as tolerating specific adversarial behaviours and staying online during the update process. NFT schemes closely rely on their underlying platforms and keep consistent with them. Although the data are often stored in separate components (such as the IPFS file system), the most important logic and tokeId are still recorded on-chain. Properly updating the system with improvements will be a necessity.

\end{itemize}

\section{Conclusion}
\label{sec-conclu}
Non-Fungible Token (NFT) is an emerging technology prevailing in the blockchain market. In this report, we explore the state-of-the-art NFT solutions which may re-shape the market of digital/virtual assets stepping forward. We firstly analyze the technical components and provide the design models and properties. Then, we evaluate the security of current NFTs systems and further discuss the opportunities and potential applications that adopt the NFT concept. Finally, we outline existing research challenges that require to be solved before achieving mass-market penetration. We hope this report delivers timely analysis and summary of existing proposed solutions and projects, making it easier for newcomers to keep up with the current progress.

\bibliographystyle{splncs04}
\bibliography{bib}

\begin{thebibliography}{100}
\providecommand{\url}[1]{\texttt{#1}}
\providecommand{\urlprefix}{URL }
\providecommand{\doi}[1]{https://doi.org/#1}

\bibitem{decentraland2020}
Decentraland (mana). Project accessible: \url{https://decentraland.org/}
  (2020)

\bibitem{flow2020}
Flow. Project accessible: \url{https://www.onflow.org/}  (2020)

\bibitem{sandbox2020}
Sandbox. Project accessible: \url{https://www.sandbox.game/en/}  (2020)

\bibitem{3dpunks}
3d punks. Project Accessible: \url{http://www.3dpunks.com/}  (2021)

\bibitem{aavegotchi}
Aavegotchi. Project Accessible: \url{https://aavegotchi.com/}  (2021)

\bibitem{alienworlds}
Alien worlds. Project Accessible: \url{https://alienworlds.io/}  (2021)

\bibitem{artblocks}
Art blocks. Project Accessible: \url{https://artblocks.io/}  (2021)

\bibitem{asyncart}
Async art. Project Accessible: \url{https://async.art/}  (2021)

\bibitem{axieinfinity}
Axie infinity. Project Accessible: \url{https://axieinfinity.com/}  (2021)

\bibitem{beeple}
Beeple. Project Accessible: \url{https://www.beeple-crap.com/}  (2021)

\bibitem{boredapeyachtclub}
Bored ape yacht club. Project Accessible:
  \url{https://boredapeyachtclub.com/#/}  (2021)

\bibitem{bullrun}
Bullrun babes. Project Accessible:
  \url{https://opensea.io/collection/bullrunbabestoken}  (2021)

\bibitem{cargo}
Cargo. Project Accessible: \url{https://cargo.build/}  (2021)

\bibitem{chubbies}
Chubbies. Project Accessible: \url{https://chubbies.io/}  (2021)

\bibitem{coingecko}
Coingecko website. Accessible: \url{https://coingecko.com/en}  (2021)

\bibitem{cometh}
Cometh. Project Accessible: \url{https://www.cometh.io/}  (2021)

\bibitem{cryptostamp}
Crypto stamp. Project Accessible: \url{https://crypto.post.at/}  (2021)

\bibitem{cryptocats}
Cryptocats. Project Accessible: \url{https://cryptocats.thetwentysix.io/}
  (2021)

\bibitem{cryptokitties}
Cryptokitties. Project Accessible: \url{https://www.cryptokitties.co/}  (2021)

\bibitem{cryptopunks}
Cryptopunks. Accessible: \url{https://www.larvalabs.com/cryptopunks}  (2021)

\bibitem{cryptoslam}
Cryptoslam. Project Accessible: \url{https://cryptoslam.io/}  (2021)

\bibitem{cryptovoxels}
Cryptovoxels. Project Accessible: \url{https://www.cryptovoxels.com/}  (2021)

\bibitem{cryptowine}
Cryptowine. Project Accessible: \url{https://grap.finance/#/}  (2021)

\bibitem{dappradar}
Dappradar website. Accessible: \url{https://dappradar.com/}  (2021)

\bibitem{dego}
Dego. Project Accessible: \url{https://dego.finance/}  (2021)

\bibitem{defipulse}
Deipulse website. Accessible: \url{https://defipulse.com/}  (2021)

\bibitem{ens}
Ens domains. Project Accessible: \url{https://app.ens.domains/}  (2021)

\bibitem{fastbox}
Fast box. Project Accessible: \url{https://www.fastbox.cc/}  (2021)

\bibitem{godsunchained}
Gods unchanged. Project Accessible: \url{https://godsunchained.com/}  (2021)

\bibitem{grimes}
Grimes sold $\$6$ million worth of digital art as nfts. News source:
  \url{https://www.theverge.com/2021/3/1/22308075/grimes-nft-6-million-sales-nifty-gateway-warnymph}
   (2021)

\bibitem{hashmasks}
Hashmasks. Project Accessible: \url{https://www.thehashmasks.com/}  (2021)

\bibitem{icecap}
Icecap. Project Accessible: \url{https://icecap.diamonds/}  (2021)

\bibitem{knownorigin}
Known origin. Project Accessible: \url{https://www.knownorigin.io/}  (2021)

\bibitem{loot}
Loot contract code.
  \url{https://etherscan.io/address/0xff9c1b15b16263c61d017ee9f65c50e4ae0113d7#code}
   (2021)

\bibitem{loottalk}
Loot talk. \url{https://loot-talk.com/}  (2021)

\bibitem{maddog}
Mad dog jones. News source:
  \url{https://www.phillips.com/detail/mad-dog-jones/NY090121/1}  (2021)

\bibitem{makersplace}
Makersplace. Project Accessible: \url{https://makersplace.com/}  (2021)

\bibitem{meebits}
Meebits. Project Accessible: \url{https://meebits.larvalabs.com/}  (2021)

\bibitem{megacryptopolis}
Megacryptopolis. Project Accessible: \url{https://mcp3d.com/}  (2021)

\bibitem{mintable}
Mintable. Project Accessible: \url{https://mintable.app/}  (2021)

\bibitem{mintbase}
Mintbase. Project Accessible: \url{https://www.mintbase.io/}  (2021)

\bibitem{mooncatsrescue}
Moon cats rescue. Project Accessible: \url{https://mooncatrescue.com/}  (2021)

\bibitem{mycryptoheroes}
Mycryptoheroes. Project Accessible: \url{https://www.mycryptoheroes.net/}
  (2021)

\bibitem{nbatopshot}
Nba top shot. Accessible: \url{https://nbatopshot.com/}  (2021)

\bibitem{nftbank}
Nft bank. Project Accessible: \url{https://nftbank.ai/}  (2021)

\bibitem{nftbox}
Nft box. Project Accessible: \url{https://nftboxes.io/}  (2021)

\bibitem{nftdiamonds}
Nft diamonds. Project Accessible: \url{https://nftdiamonds.co/}  (2021)

\bibitem{niftygateway}
Nifty gateway. Project Accessible: \url{https://niftygateway.com/}  (2021)

\bibitem{NonFungible}
Nonfungible website. Accessible: \url{https://NonFungible.com}  (2021)

\bibitem{opensea}
Opensea platform. Accessible: \url{https://opensea.io/}  (2021)

\bibitem{picassopunks}
Picasso punks. Accessible: \url{https://opensea.io/collection/picassopunks}
  (2021)

\bibitem{polkamon}
Polkamon. Project Accessible: \url{https://polkamon.com/}  (2021)

\bibitem{rplanet}
R planet. Project Accessible: \url{https://rplanet.io/}  (2021)

\bibitem{lagelnd}
R.a.r.e art lab. Project Accessible: \url{https://www.lagelnd.com/rare}  (2021)

\bibitem{rarible}
Rarible. Project Accessible: \url{https://rarible.com/}  (2021)

\bibitem{skyweaver}
Skyweaver. Project Accessible: \url{https://www.skyweaver.net/}  (2021)

\bibitem{somniumspace}
Somnium space. Project Accessible: \url{https://somniumspace.com/}  (2021)

\bibitem{sorare}
Sorare. Project Accessible: \url{https://sorare.com/}  (2021)

\bibitem{superrare}
Superrare. Project Accessible: \url{https://superrare.co/}  (2021)

\bibitem{toppsmlb}
Topps mlb. Project Accessible: \url{https://toppsmlb.com/}  (2021)

\bibitem{tradestars}
Tradestars. Project Accessible: \url{https://tradestars.app/}  (2021)

\bibitem{trevorjonesart}
Trevorjonesart. Project Accessible: \url{https://www.trevorjonesart.com/}
  (2021)

\bibitem{unofficialpunks}
Unofficial punks. Project Accessible:
  \url{https://opensea.io/collection/unofficialpunks}  (2021)

\bibitem{unstoppabledomains}
Unstoppable domains. Accessible: \url{https://unstoppabledomains.com/}  (2021)

\bibitem{viv3}
Viv3. Project Accessible: \url{https://VIV3.com}  (2021)

\bibitem{wax2}
Wax: Worldwide asset exchange. Accessible: \url{https://on.wax.io/wax-io/}
  (2021)

\bibitem{wax}
Wax:worldwide asset exchange. Project Accessible:
  \url{https://github.com/worldwide-asset-exchange/whitepaper}  (2021)

\bibitem{wiv}
Wiv. Project Accessible: \url{https://www.wiv.io/}  (2021)

\bibitem{zora}
Zora. Project Accessible: \url{https://zora.co/}  (2021)

\bibitem{bal2019nftracer}
Bal, M., Ner, C.: Nftracer: a non-fungible token tracking proof-of-concept
  using hyperledger fabric. arXiv preprint arXiv:1905.04795  (2019)

\bibitem{bano2019sok}
Bano, S., Sonnino, A., Al-Bassam, M., Azouvi, S., McCorry, P., Meiklejohn, S.,
  Danezis, G.: Sok: Consensus in the age of blockchains. In: Proceedings of the
  1st ACM Conference on Advances in Financial Technologies. pp. 183--198 (2019)

\bibitem{benet2014ipfs}
Benet, J.: Ipfs-content addressed, versioned, p2p file system. arXiv preprint
  arXiv:1407.3561  (2014)

\bibitem{YesYourN79}
Benson, J.: Your nfts can go missing—here's what you can do about it.
  \url{https://decrypt.co/62037/missing-or-stolen-nfts-how-to-protect} (2021)

\bibitem{buterin2014next}
Buterin, V., et~al.: A next-generation smart contract and decentralized
  application platform. white paper  \textbf{3}(37) (2014)

\bibitem{cai2018decentralized}
Cai, W., Wang, Z., et~al.: Decentralized applications: The blockchain-empowered
  software system. IEEE Access  \textbf{6},  53019--53033 (2018)

\bibitem{CH21}
CH21: Check my nft. \url{https://checkmynft.com/} (2021)

\bibitem{chevet2018blockchain}
Chevet, S.: Blockchain technology and non-fungible tokens: Reshaping value
  chains in creative industries. Available at SSRN 3212662  (2018)

\bibitem{chohan2021non}
Chohan, U.W.: Non-fungible tokens: Blockchains, scarcity, and value. Critical
  Blockchain Research Initiative (CBRI) Working Papers  (2021)

\bibitem{ciampi2020updatable}
Ciampi, M., et~al.: Updatable blockchains. In: European Symposium on Research
  in Computer Security. pp. 590--609. Springer (2020)

\bibitem{dowling2021fertile}
Dowling, M.M.: Fertile land: Pricing non-fungible tokens. Available at SSRN
  3813522  (2021)

\bibitem{erc20}
Fabian, V., Vitalik, B.: Eip-20: Erc-20 token standard. Accessible:
  \url{https://eips.ethereum.org/EIPS/eip-20}  (2015)

\bibitem{fairfield2021tokenized}
Fairfield, J.: Tokenized: The law of non-fungible tokens and unique digital
  property. Indiana Law Journal, Forthcoming  (2021)

\bibitem{franceschet2020crypto}
Franceschet, M., Colavizza, G., Smith, T., et~al.: Crypto art: A decentralized
  view. Leonardo pp.~1--8 (2020)

\bibitem{garay2020sok}
Garay, J., Kiayias, A.: Sok: A consensus taxonomy in the blockchain era. In:
  Cryptographers’ Track at the RSA Conference. pp. 284--318. Springer (2020)

\bibitem{garay2015bitcoin}
Garay, J., Kiayias, A., Leonardos, N.: The bitcoin backbone protocol: Analysis
  and applications. In: Annual International Conference on the Theory and
  Applications of Cryptographic Techniques (EUROCRYPT). pp. 281--310. Springer
  (2015)

\bibitem{garay2017bitcoin}
Garay, J., Kiayias, A., Leonardos, N.: The bitcoin backbone protocol with
  chains of variable difficulty. In: CRYPTO. pp. 291--323. Springer (2017)

\bibitem{gervais2016security}
Gervais, A., et~al.: On the security and performance of proof of work
  blockchains. In: Proceedings of the 2016 ACM SIGSAC Conference on Computer
  and Communications Security. pp. 3--16. ACM (2016)

\bibitem{gudgeon2020sok}
Gudgeon, L., Moreno-Sanchez, P., Roos, S., McCorry, P., Gervais, A.: Sok:
  Layer-two blockchain protocols. In: International Conference on Financial
  Cryptography and Data Security. pp. 201--226. Springer (2020)

\bibitem{hong2019design}
Hong, S., Noh, Y., Park, C.: Design of extensible non-fungible token model in
  hyperledger fabric. In: Proceedings of the 3rd Workshop on Scalable and
  Resilient Infrastructures for Distributed Ledgers. pp.~1--2 (2019)

\bibitem{erc777}
Jacques, D., Jordi, B., Thomas, S.: Eip-777: Erc-777 token standard.
  Accessible: \url{https://eips.ethereum.org/EIPS/eip-777}  (2017)

\bibitem{johnson2021decentralized}
Johnson, K.N.: Decentralized finance: Regulating cryptocurrency exchanges.
  William \& Mary Law Review  \textbf{62} (2021)

\bibitem{lamport2019byzantine}
Lamport, L., Shostak, R., Pease, M.: The byzantine generals problem. In:
  Concurrency: the Works of Leslie Lamport, pp. 203--226 (2019)

\bibitem{li2019auditable}
Li, R., Galindo, D., Wang, Q.: Auditable credential anonymity revocation based
  on privacy-preserving smart contracts. In: Data Privacy Management,
  Cryptocurrencies and Blockchain Technology, pp. 355--371. Springer (2019)

\bibitem{li2020accountable}
Li, R., et~al.: An accountable decryption system based on privacy-preserving
  smart contracts. In: International Conference on Information Security. pp.
  372--390. Springer (2020)

\bibitem{menezes2018handbook}
Menezes, A.J., Van~Oorschot, P.C., Vanstone, S.A.: Handbook of applied
  cryptography. CRC press (2018)

\bibitem{Moore2006InferringID}
Moore, D., Voelker, G., Savage, S.: Inferring internet denial-of-service
  activity. ACM Trans. Comput. Syst.  \textbf{24},  115--139 (2006)

\bibitem{musan2020nft}
Musan, D.I., William, J., Gervais, A.: Nft. finance leveraging non-fungible
  tokens  (2020)

\bibitem{nakamoto2019bitcoin}
Nakamoto, S.: Bitcoin: A peer-to-peer electronic cash system. Tech. rep.,
  Manubot (2019)

\bibitem{noether2015ring}
Noether, S.: Ring signature confidential transactions for monero. IACR Cryptol.
  ePrint Arch.  \textbf{2015}, ~1098 (2015)

\bibitem{omar2020capability}
Omar, A.S., Basir, O.: Capability-based non-fungible tokens approach for a
  decentralized aaa framework in iot. In: Blockchain Cybersecurity, Trust and
  Privacy, pp. 7--31. Springer (2020)

\bibitem{raman2018trusted}
Raman, R.K., et~al.: Trusted multi-party computation and verifiable
  simulations: A scalable blockchain approach. arXiv preprint arXiv:1809.08438
  (2018)

\bibitem{raval2016decentralized}
Raval, S.: Decentralized applications: harnessing Bitcoin's blockchain
  technology. " O'Reilly Media, Inc." (2016)

\bibitem{regner2019nfts}
Regner, F., Urbach, N., Schweizer, A.: Nfts in practice--non-fungible tokens as
  core component of a blockchain-based event ticketing application  (2019)

\bibitem{rogaway2004cryptographic}
Rogaway, P., Shrimpton, T.: Cryptographic hash-function basics: Definitions,
  implications, and separations for preimage resistance, second-preimage
  resistance, and collision resistance. In: FSE. pp. 371--388. Springer (2004)

\bibitem{shirole2020cryptocurrency}
Shirole, M., Darisi, M., Bhirud, S.: Cryptocurrency token: An overview. IC-BCT
  2019 pp. 133--140 (2020)

\bibitem{shostack2008experiences}
Shostack, A.: Experiences threat modeling at microsoft. MODSEC@ MoDELS
  \textbf{2008} (2008)

\bibitem{szabo1996smart}
Szabo, N.: Smart contracts: building blocks for digital markets. EXTROPY: The
  Journal of Transhumanist Thought,(16)  \textbf{18}(2) (1996)

\bibitem{trautman2021virtual}
Trautman, L.J.: Virtual art and non-fungible tokens. Available at SSRN 3814087
  (2021)

\bibitem{valdeolmillos2019blockchain}
Valdeolmillos, D., et~al.: Blockchain technology: a review of the current
  challenges of cryptocurrency. In: International Congress on Blockchain and
  Applications. pp. 153--160. Springer (2019)

\bibitem{wang2019sok}
Wang, G., et~al.: Sok: Sharding on blockchain. In: Proceedings of the 1st ACM
  Conference on Advances in Financial Technologies. pp. 41--61 (2019)

\bibitem{wang2021weak}
Wang, Q., Li, R.: A weak consensus algorithm and its application to
  high-performance blockchain. In: IEEE INFOCOM 2021-IEEE Conference on
  Computer Communications (INFOCOM). IEEE (2021)

\bibitem{wang2020preserving}
Wang, Q., Qin, B., Hu, J., Xiao, F.: Preserving transaction privacy in bitcoin.
  Future Generation Computer Systems  \textbf{107},  793--804 (2020)

\bibitem{wang2020sok}
Wang, Q., Yu, J., Chen, S., Xiang, Y.: Sok: Diving into dag-based blockchain
  systems. arXiv preprint arXiv:2012.06128  (2020)

\bibitem{wang2018designing}
Wang, Y., Kogan, A.: Designing confidentiality-preserving blockchain-based
  transaction processing systems. International Journal of Accounting
  Information Systems  \textbf{30},  1--18 (2018)

\bibitem{erc721}
William, E., Dieter, S., Jacob, E., Nastassia, S.: Eip-721: Erc-721
  non-fungible token standard. Accessible:
  \url{https://eips.ethereum.org/EIPS/eip-721}  (2018)

\bibitem{eip721}
William, E., Dieter, S., Jacob, E., Nastassia, S.: Erc-721 non-fungible token
  standard. Ethereum Improvement Protocol, EIP-721, Accessible:
  \url{https://eips.ethereum.org/EIPS/eip-721}.  (2018)

\bibitem{eip1155}
Witek, R., Andrew, C., Philippe, C., James, T., Eric, B., Ronan, S.: Eip-1155:
  Erc-1155 multi token standard. Ethereum Improvement Protocol, EIP-1155,
  Accessible: \url{https://eips.ethereum.org/EIPS/eip-1155}.  (2018)

\bibitem{erc1155}
Witek, R., et~al.: Eip-1155: Erc-1155 multi token standard. Accessible:
  \url{https://eips.ethereum.org/EIPS/eip-1155}  (2018)

\bibitem{wood2014ethereum}
Wood, G., et~al.: Ethereum: A secure decentralised generalised transaction
  ledger. Ethereum project yellow paper  \textbf{151}(2014),  1--32 (2014)

\bibitem{zamyatin2019sok}
Zamyatin, A., et~al.: Sok: communication across distributed ledgers.  (2019)

\bibitem{zhou1996fair}
Zhou, J., Gollman, D.: A fair non-repudiation protocol. In: Proceedings 1996
  IEEE Symposium on Security and Privacy. pp. 55--61. IEEE (1996)

\end{thebibliography}

\section*{Appendix A. Version Control}

This work is an ongoing updated technique report. In the future, we will add the landmark events from \textit{in the wild} NFT projects to inspire technique movements and developments. Dynamic data (like prices, sales, or market cap) will remain as it is in the current snapshot [as of May 2021].

\begin{table}[!hbtp]
 \caption{Version Updates} 
 \label{tab-nftrank}
  \centering
 \resizebox{\linewidth}{!}{
 \begin{tabular}{crl }
    \toprule
    \textbf{Version} & \multicolumn{1}{c}{\textbf{Date}} & \multicolumn{1}{c}{\textbf{Updates}}  \\
    \midrule
     
   V1  & May 2021  & \quad Main body construction \\
   V2 & October 2021 & \quad Adding the \textit{bottom to top} model in Section 3  \\ 
     & & \quad Adding the instance analysis of the Loot project in Appendix D \\
     & & \quad Revisions of unclear sentences and paragraphs \\
    \bottomrule
  \end{tabular}
 }
\end{table}

\section*{Appendix B. Top Sales of NFT Properties}
\label{Appendixa}

\begin{table}[!hbtp]
 \caption{ NFT Collectible Ranking by Sales Volume (All-time)} 
 \label{tab-nftrank}
  \centering
 \resizebox{\linewidth}{!}{
 \begin{tabular}{ll p{3cm}p{2.2cm}p{2.2cm}p{2.2cm}}
    \toprule
    \textbf{Rank} &\textbf{Product} & \textbf{Sales}  & \textbf{Buyers} &
    \textbf{Txns} & \textbf{Owners} \\
    \midrule
     
   1  & NBA Top Shot \cite{nbatopshot}  & $\$$573,815,060.38  &  285,306 & 4,877,975 & 498,659 \\
   2 & CryptoPunks \cite{cryptopunks} &  $\$$316,956,115.15  &   2,716 & 13,231 & 2,282\\
   3 & Meebits \cite{meebits}  &  $\$$54,750,807.84 &  1,336 & 2,995 & 4,436\\
   4 & Hashmasks \cite{hashmasks}  & $\$$49,713,202.76  &  3,165 & 11,322 & 4,254 \\
   5 & Sorare \cite{sorare}  &  $\$$39,807,257.19 & 16,435 & 225,909 & 18,050\\
   6 & CryptoKitties \cite{cryptokitties}  &  $\$$33,230,422.66 & 100,624 & 762,562 & \\
   7 & Art Blocks \cite{artblocks}  & $\$$18,824,119.18  & 2,052 & 10,907 & 4,521 \\
   8 & Alien Worlds \cite{alienworlds}  & $\$$17,239,575.21  & 165,818 &	2,280,968 &	1,124,901 \\
   9 & Bored Ape Yacht Club \cite{boredapeyachtclub}  & $\$$11,844,235.71  & 2,281 & 5,786 & 2,697 \\
   10 & Topps MLB \cite{toppsmlb}  &  $\$$10,205,104.07 & 12,706 & 365,963 & 36,283 \\

    \bottomrule
  \end{tabular}
  }
  \begin{tablenotes}
       \footnotesize
       \item[1] \quad Data captured on May 10, 20201, source from: \url{https://cryptoslam.io/} 
   \end{tablenotes}
\end{table}

\newpage
\section*{Appendix C. Overview of Existing NFT solutions}

\begin{table}[!hbtp]
 \caption{Existing NFTs Projects (May, 2021)} 
 \label{tab-nftprojec}
 \begin{threeparttable}
  \centering
 \resizebox{\linewidth}{!}{
 \begin{tabular}{llll}
    \toprule
    \textbf{Types} & \textbf{Projects} & \textbf{Platform} \quad &
    \quad\quad\quad\quad \textbf{Key Words} \\
    \midrule
    \multirow{4}{*}{\textit{Collectible}}  
    & NBA Top Shot \cite{nbatopshot}  & Flow \cite{flow2020} & Basketball moments \\
    & CryptoPunks \cite{cryptopunks} &  Ethereum  & The first NFT products \\
    & CryptoWine \cite{cryptowine}  &  Ethereum & Mining by graps / Trading wines \\
    & Hashmask \cite{hashmasks}  &  Ethereum & Digital arts / Blind boxes  \\
    \cmidrule{1-2}
    
    \multirow{6}{*}{\textit{Cards}} 
    & Sorare \cite{sorare} & Ethereum & Football star cards \\
    & Skyweaver \cite{skyweaver} & Ethereum & Card game \\
    & Cometh \cite{cometh} & Ethereum & Galaxy background \\
    & Gods Unchained \cite{godsunchained} & Ethereum & Trading card game / Hearthstone\\
    & Topps MLB \cite{toppsmlb} & WAX \cite{wax2}& Baseball cards \\
    & R Planet \cite{rplanet} & WAX & Red planet background  \\
    \cmidrule{1-2}
    
    \multirow{3}{*}{\textit{Raising Pets}} 
    & CryptoKitties \cite{cryptokitties} &  Ethereum & The first game with hype in Ethereum \\
     &  CryptoCats \cite{cryptocats} & Ethereum & \multirow{2}{*}{ Raising pets and sell with high prices}  \\
     & Axie Infinity \cite{axieinfinity} & Ethereum &  \\
    \cmidrule{1-2}
    
    \multirow{6}{*}{\textit{Virtual World}} & Sandbox \cite{sandbox2020} & Ethereum & \multirow{5}{*}{ \makecell[l]{Create, explore and trade in the virtual \\ world owned by users, also make money \\by selling the properties.}} \\
     & Decentraland \cite{decentraland2020} &  Ethereum &  \\
     & Sommnium Space \cite{somniumspace} &  Ethereum &   \\
     & Cryptovoxels \cite{cryptovoxels} &  Ethereum &  \\
     & Alien Worlds \cite{alienworlds} & WAX  & \\
     & MyCryptoHeros \cite{mycryptoheroes} &  Ethereum & Historical story / Japan made \\
    \cmidrule{1-2}
    
    \multirow{4}{*}{\textit{Real Properties}} 
     & Crypto stamp \cite{cryptostamp} & Ethereum & NFTs on stamps \\
     & Diamonds \cite{nftdiamonds} & Ethereum &\multirow{2}{*}{NFTs on diamonds} \\
     & Icecap \cite{icecap} & Ethereum & \\
     & Wiv \cite{wiv} & Ethereum &  NFTs on wine \\
     \cmidrule{1-2}
    
    \multirow{6}{*}{\textit{Art Market}} 
    & SuperRare \cite{superrare}  & \quad\quad - &  \multirow{6}{*}{\makecell[l]{A marketplace to create, collect and  \\ trade unique, single-edition artworks. \\Every artwork is authentically created \\ by artists, and tokenized as a crypto-\\collectible  digital item.}} \\
    & Rarible \cite{rarible}  & \quad\quad - & \\
    & Cargo \cite{cargo}  & \quad\quad - & \\
    & Async Art \cite{asyncart}  & \quad\quad - & \\
    & Nifty Gateway \cite{niftygateway}  & \quad\quad - & \\
    & KnownOrigin \cite{knownorigin}  & \quad\quad - & \\
    \cmidrule{1-2}
     
     \multirow{5}{*}{Statistic Web} 
     & NonFungible \cite{NonFungible} &\quad\quad -&  \multirow{5}{*}{\makecell[l]{ Trace and monitor NFT-related \\ projects or generic types of \\cryptocurrencies by providing  timely \\ statistical data to show trends.}} \\
     & DappRadar \cite{dappradar} &\quad\quad -& \\
     & NFT bank\cite{nftbank} &\quad\quad -& \\
     & DefiPulse \cite{defipulse} &\quad\quad -& \\
     & Coingecko \cite{coingecko} &\quad\quad -& \\
     \cmidrule{1-2}
     
     \multirow{3}{*}{Exchanges} 
     & Cryptoslam \cite{cryptoslam} && \multirow{3}{*}{\makecell[l]{A marketplace to trade unique, \\single-edition NFT-based properties.}} \\
     & Opensea \cite{opensea} & \quad\quad - & \\
     & Zora \cite{zora}) &\quad\quad - & \\
    
    \bottomrule
  \end{tabular}
  }
    \begin{tablenotes}
       \footnotesize
       \item[] \quad Sources from NonFungible, Cryptoslam, DefiPulse, Coingecko, and DappRadar.
   \end{tablenotes}
  \end{threeparttable}
\end{table}

\newpage

\section*{Appendix D. Analysis of Loot}
  
Loot \cite{loot}\cite{loottalk} is a type of NFT product that uses ERC protocol on Ethererum, with totally $8,000$ entries ($7778$ for trading, $222$ for team incentives). The project presents all their products in the form of TEXT, removing functionalities of images, stats, or any representative formats. Users can only see eight lines of sentences, describing the attributes of game gears (e.g., weapons, armor, necklaces, Rings). In order to create scarcity,  Loot adds randomized prefixes (with $42\%$ probability) or suffixes ($8.7\%$) for each line of attributes (see the following code, line 2), according to the assigned tokenID from users. Users can only use Etherscan's smart contract to cast loot, rather than its official website. 


\begin{lstlisting}[language=Matlab,
                   stepnumber=1,
                   numbers=left,
                   basicstyle=\scriptsize\ttfamily,
                   numbersep=5pt,
                   tabsize=2,
                   showspaces=false,
                   frame=single,
                   title={Selected code in Loot project},
                   showstringspaces=false]
function pluck(tokenId, keyPrefix, sourceArray) internal view returns (string memory) {
    uint256 rand = random(string(abi.encodePacked(keyPrefix, toString(tokenId))));
    string memory output = sourceArray[rand % sourceArray.length];
    uint256 greatness = rand % 21;
    if (greatness > 14) {
        output = string(abi.encodePacked(output, " ", suffixes[rand % suffixes.length]));
    }
    if (greatness >= 19) {
        string[2] memory name;
        name[0] = namePrefixes[rand % namePrefixes.length];
        name[1] = nameSuffixes[rand % nameSuffixes.length];
        if (greatness == 19) {
            output = string(abi.encodePacked('"', name[0], ' ', name[1], '" ', output));
        } else {
            output = string(abi.encodePacked('"', name[0], ' ', name[1], '" ', output, " +1"));
        }
    }
    return output;
}
\end{lstlisting}
As faced by other NFTs, Loot confronts many drawbacks either. Firstly, the value of NFT is uncertain. Loot is still in the early stage of development in the current stage. Despite the price continuing to rise, there is no lack of hype.  The real consensus from communities should be maintained for a long period. Secondly, difficulty in minting and high gas cost retard its wide participant. Loot does not have a front-end Dapps, which can only be minted in the Web3 browser. Complex steps become a big challenge for newcomers. Meanwhile, the high gas fee of operations in Ethereum keeps many people away from this game. The trading or minting fees can reach up to tens of hundreds of dollars for each operation. Thirdly,  the extensibility needs further improvements. The current version (experimental stage) merely covers eight explicit attributes in the RPG game. Also, there is no space for on-top applications. The only future might be to develop more attributes and slots for sale.

\end{document}